\documentclass[conference]{IEEEtran}

\usepackage{amsfonts, amsmath, amsthm, amssymb}
\usepackage{mathtools, graphicx, epstopdf, tikz, standalone}
\usepackage{algorithm}
\usepackage[noend]{algpseudocode}
\usepackage{subcaption} 
\newcommand{\h}{\hat{V_s}}
\newcommand{\w}{\omega}
\newcommand{\von}{V_{\rm on}}
\newcommand{\vout}{V_{\rm out}}
\newcommand{\ron}{R_{\rm on}}
\newcommand{\roff}{R_{\rm off}}
\newcommand{\Ton}{T_{\rm on}}
\newcommand{\Toff}{T_{\rm off}}
\newcommand{\ton}{t_0^{\rm on}}
\newcommand{\toff}{t_0^{\rm off}}

\IEEEoverridecommandlockouts
\allowdisplaybreaks

\begin{document}
\title{Integrated SWIPT Receiver with Memory Effects:\\Circuit Analysis and Information Detection}	
\author{
\IEEEauthorblockN{Eleni Demarchou\IEEEauthorrefmark{2}, Zulqarnain Bin Ashraf\IEEEauthorrefmark{1}, Dieff Vital\IEEEauthorrefmark{1},\\ Besma Smida\IEEEauthorrefmark{1}, Constantinos Psomas\IEEEauthorrefmark{2}, and Ioannis Krikidis\IEEEauthorrefmark{2}}\vspace{0.1mm}
\IEEEauthorblockA{\IEEEauthorrefmark{2}Department of Electrical and Computer Engineering, University of Cyprus, Cyprus}\vspace{0.1mm}
\IEEEauthorblockA{\IEEEauthorrefmark{1}Department of Electrical and Computer Engineering, University of Illinois at Chicago, USA}
\IEEEauthorblockA{e-mail: edemar01@ucy.ac.cy, zbinas2@uic.edu, dvital@uic.edu, smida@uic.edu, psomas@ucy.ac.cy, krikidis@ucy.ac.cy}
\thanks{This work received funding from the European Research Council (ERC) under the European Union's Horizon 2020 research and innovation programme (Grant agreement No. 819819). It was also funded by the European Union's Horizon Europe programme (ERC, WAVE, Grant agreement No. 101112697), and from the European Union HORIZON programme under iSEE-6G GA No. 101139291.}
\vspace*{-9mm}}
\maketitle

\begin{abstract}
Wireless power transfer has been proposed as a key technology for the foreseen machine type networks. A main challenge in the research community lies in acquiring a simple yet accurate model to capture the energy harvesting performance. In this work, we focus on a half-wave rectifier and based on circuit analysis we provide the actual output of the circuit which accounts for the memory introduced by the capacitor. The provided expressions are also validated through circuit simulations on ADS. Then, the half-wave rectifier is used as an integrated simultaneous wireless information and power transfer receiver where the circuit's output is used for decoding information based on amplitude modulation.
We investigate the bit error rate performance based on two detection schemes: (i) symbol-by-symbol maximum likelihood (ML); and (ii) ML sequence detection (MLSD). We show that the symbol period is critical due to the intersymbol interference induced by circuit. Our results reveal that MLSD is necessary towards improving the error probability and achieving higher data rates. 
\end{abstract}

\begin{IEEEkeywords}
wireless power transfer, half-wave rectifier, circuit analysis, integrated SWIPT receiver.
\end{IEEEkeywords}

\section{Introduction}
Wireless power transfer (WPT) has been proposed as a key technology for the sixth generation (6G) networks, as a flexible and viable solution for powering up the low demanding devices of the foreseen machine type networks \cite{Bennis}. This is achieved by integrating into a radio frequency (RF) antenna a rectifying circuit which converts the ambient or dedicated electromagnetic radiation into direct current (DC) \cite{thesis}. A conventional rectifier consists of a single diode and a capacitor which flattens the high output oscillations. This topology is known as half-wave rectifier and it provides high RF to DC conversion efficiency at low input power \cite{ThesisMorsi}.

The benefits provided by WPT have attracted the interest of the research community where a critical challenge lies in applying a valid model for capturing the energy harvesting (EH) performance of the rectifier circuit. The linear input-output model has been widely used in the past years \cite{linear}, but it only takes into account the circuit's conversion efficiency while it is independent of the input power level and neglects the non-linear behavior of the rectifier's components. A more practical approach is presented in \cite{sigmoidal}, where the authors propose a non-linear EH model based on the sigmoidal function, capturing the dynamics of the conversion efficiency for different input power levels. Moreover, the authors in \cite{Bruno}, present an analytical model of the rectenna non-linearity through the Taylor expansion of the diode characteristics. Another model is proposed in \cite{Morsi}, where the authors take into account both the forward and reverse current of the rectifying diode as well as the matching network and provide an expression for the harvested DC power. Furthermore, the authors in \cite{Schober}, take into account the memory introduced to the system due to the capacitor and model the circuit through a Markov decision process. The state transitions probabilities are obtained through a learning based approach by taking into account all the non-linear effects of the rectifier as well as the impedance mismatch.

Besides, RF signals constitute the fundamental medium for wireless communications and their ability to transfer energy has leveraged a significant interest in simultaneous wireless information and power transfer (SWIPT) systems \cite{Varshney}. The authors in \cite{Kim}, consider the SWIPT receiver utilizing a time-switching method to separate information decoding (ID) and EH modes. Another, well-investigated, approach for employing SWIPT systems, is the power-splitting protocol where a portion of the received signal is used for ID and the remaining part is used for EH \cite{Poor}. The aforementioned SWIPT receivers implementing the time or power splitting approach, fall under the category of separated circuit architecture where the received signal is split towards EH and ID at the RF antenna front end. Another category, is the integrated SWIPT receiver, first introduced in \cite{RuiZ}, where the authors suggest to split the signal towards EH and ID at the rectifier's end, avoiding the need for RF to baseband conversion and thereafter reducing the circuit complexity of the SWIPT receiver. A more compact SWIPT solution is introduced by the authors in \cite{SofieC}, suggesting to rely on amplitude modulation and exploit the rectifier's circuit to jointly design the EH and ID. Specifically, the authors suggest to apply a biased amplitude shift keying (ASK) where the minimum amplitude is non-zero in order to avoid zero harvested power. The authors extended their work in \cite{SofieJ}, where they derive an analytical symbol error rate expression for the integrated receiver circuit in the presence of additive white Gaussian noise (AWGN). However, the analytical circuit output is not provided while the memory effects of the system are not investigated.

Different from previous works, in this paper, we take into account the memory induced by the capacitor and provide a simple yet accurate model to characterize the output voltage. The model, based on a half-wave rectifier, is derived through circuit analysis and its validity is confirmed with Advanced Design System (ADS) simulations. The half-wave rectifier model is then explored in the context of a SWIPT scenario. In particular, we consider an amplitude modulation scheme through which an integrated SWIPT receiver can decode symbols based on the circuit's output. We focus on the achieved bit error rate (BER) in the presence of noise but also on the RF EH. We investigate two detection schemes: (i) symbol-by-symbol maximum likelihood (ML); and (ii) ML sequence detection (MLSD). We show that, with ML detection, the symbol period is critical to the performance since intersymbol interference arises due to the capacitor's memory effects. Therefore, we apply MLSD, which improves the BER and allows for higher data rates. It is also shown that the MLSD converges to the conventional ML when the memory of the circuit is limited.

\section{System Model}
\begin{figure}\centering
\includegraphics[width=0.6\linewidth]{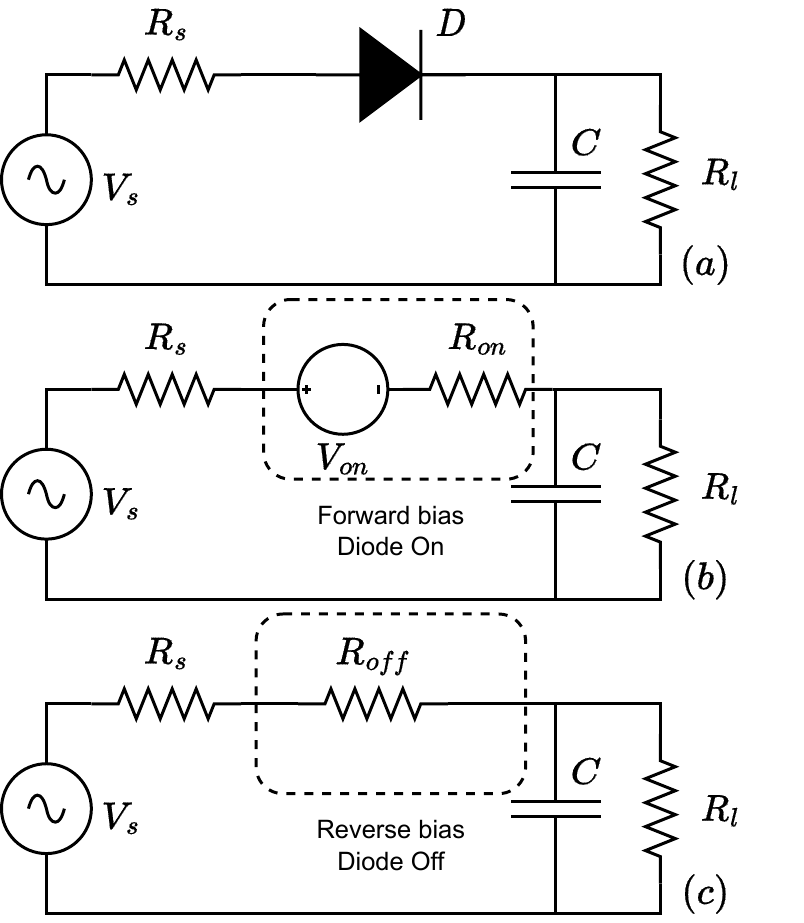}
\caption{(a) Half-wave rectifier and equivalent circuits when the diode is on (b) and off (c).}\label{circuit}
\vspace{-3mm}
\end{figure}

Consider a wireless communications antenna which harvests energy through a half-wave rectifying circuit. The circuit is composed of a single diode $D$, enabling the rectification, a capacitor $C$, for flattening high output oscillations, and a load $R_l$. The received signal at the antenna which excites the circuit is modeled as a sinusoidal source with amplitude $\h$, expressed by
\begin{equation}
V_s(t) = \h \sin\left(\w t\right),
\end{equation}
where $\w$ denotes the radial frequency. The antenna's resistance $R_s$, is also taken into account, and it is connected to the equivalent circuit as shown in Fig. \ref{circuit}(a).

For the diode's current-voltage characteristic we adopt a piece-wise model \cite{thesis}. Let $V_D(t)$ and $I_D(t)$ denote the voltage drop and current at the diode, then
\begin{equation}
I_D(t)=\begin{cases}
\frac{V_D(t)-\von}{\ron} \text{ if } V_D(t) \geq \von,\\
\frac{V_D(t)}{\roff} \text{ if } V_D(t) < \von,
\end{cases}
\end{equation}
where $\von$ denotes the minimum voltage required for the diode to be switched on, and $\ron$ and $\roff$ correspond to the diode's resistance when switched on and off, respectively. Note that, the minimum threshold $\von$ is modeled as a DC component in the equivalent circuit. This piece-wise model enables the circuit analysis to be conducted for the individual states of the diode being on and off \cite{thesis}. The equivalent circuits are depicted in Fig.  \ref{circuit}.

In what follows, we first obtain the output voltage of the rectifier and then based on the circuit characteristics we investigate an amplitude modulation scheme such that the rectifier can be used as an integrated SWIPT receiver.

\section{Half-Wave Rectifier's Output}
In this section, we provide the half-wave rectifier's output voltage as a function of the antenna's input by making use of the piece-wise model for the diode. For this purpose, we first provide the circuit analysis for each of the cases where the diode is forward or reverse biased by using the equivalent circuits depicted in Fig. \ref{circuit}(b) and (c), respectively. Then, we provide the algorithm that is used to obtain the circuit's output voltage.
\subsection{Forward biased diode}\label{sON}
We first focus on the case where the diode is on, and the equivalent circuit which is depicted in Fig. \ref{circuit}(b). By applying the Kirchoff's law for voltage we get
\begin{equation}\label{Kv1}
V_s(t)-(R_s + \ron)I_1(t)-\von-V_C(t)=0,
\end{equation}
where $V_C(t)$ denotes the output voltage i.e., the capacitor's voltage, and $I_1(t)$ is given by
\begin{align}\label{Kc1}
I_1(t)&=I_C(t)+I_{R_l}(t)\nonumber\\
&=C\frac{dV_C(t)}{dt}+\frac{V_C(t)}{R_l}.
\end{align}
Let $\Ton \triangleq C(R_s+\ron)$ and $\alpha\triangleq\frac{1}{\Ton}+\frac{1}{R_lC}$, then from the expressions in \eqref{Kv1} and \eqref{Kc1} we get
\begin{align}\label{dif1}
\frac{dV_C(t)}{dt}+\alpha V_C(t) = \frac{V_s(t)-\von}{\Ton}.
\end{align}
By considering that the diode switched on at $\ton < t$ with $V_C(\ton) = V_0^{\rm on}$, the solution to the differential equation in \eqref{dif1} is given by
\begin{align}\label{Don}
&V_C(t)=-\frac{\von}{\alpha\Ton} -\frac{\hat{V}_s(\omega\cos(\omega t) - \alpha \sin(\omega t))}{\Ton(\alpha^2+\omega^2)}\nonumber\\
&+\! e^{\alpha(\ton-t)}\bigg(\!V_0^{\rm on}+\frac{\von}{\alpha \Ton} + \frac{\hat{V}_s (\omega \cos(\omega \ton)\!-\!\alpha \sin (\omega \ton))}{\Ton (\alpha^2+\omega^2)}\!\bigg).
\end{align}
Note that, eq. \eqref{Don} defines the rectifier's output up until the following expression becomes true
\begin{equation}
\frac{(V_s(t)-V_C(t)-\von)\ron}{R_s+\ron}<0,
\end{equation}
satisfying the condition $V_D(t)<\von$.
\addtolength{\topmargin}{+0.126cm}
\subsection{Reverse biased diode}
We now turn our attention to the case where the diode is switched off and modeled as a high resistance  in the equivalent circuit as depicted in Fig. \refeq{circuit}(c). By following a similar procedure as in Section \ref{sON}, we get the following differential equation
\begin{align}
\frac{dV_C(t)}{dt}+\beta V_C(t)=\frac{V_s(t)}{\Toff},
\end{align}
where $\Toff = C(R_s+\roff)$ and $\beta=\frac{1}{\Toff}+\frac{1}{R_lC}$. By considering that the diode switched off at $\toff < t$, with $V_C(\toff) = V_0^{off}$, the rectifier's output voltage is given by
\begin{align}\label{Doff}
&V_C(t)=-\frac{\hat{V}_s (\omega\cos(\omega t)-\beta \sin(\omega t))}{\Toff(\beta^2+\omega^2)}\nonumber\\
&+e^{\beta(\toff-t)}\bigg(V_0^{\rm off} + \frac{\hat{V}_s (\omega \cos(\omega \toff)-\beta \sin (\omega \toff))}{\Toff(\beta^2+\omega^2)}\bigg).
\end{align}
The expression in \eqref{Doff} defines the output voltage until the diode switches on i.e., when
\begin{equation}
(V_s(t)-V_C(t))\frac{\roff}{R_s+\roff}\geq \von,
\end{equation}
satisfying the condition $V_D(t)\geq \von$.
\subsection{Transient Time Analysis}\label{TTA}
\begin{figure}[t]\centering
\includegraphics[width=0.9\linewidth]{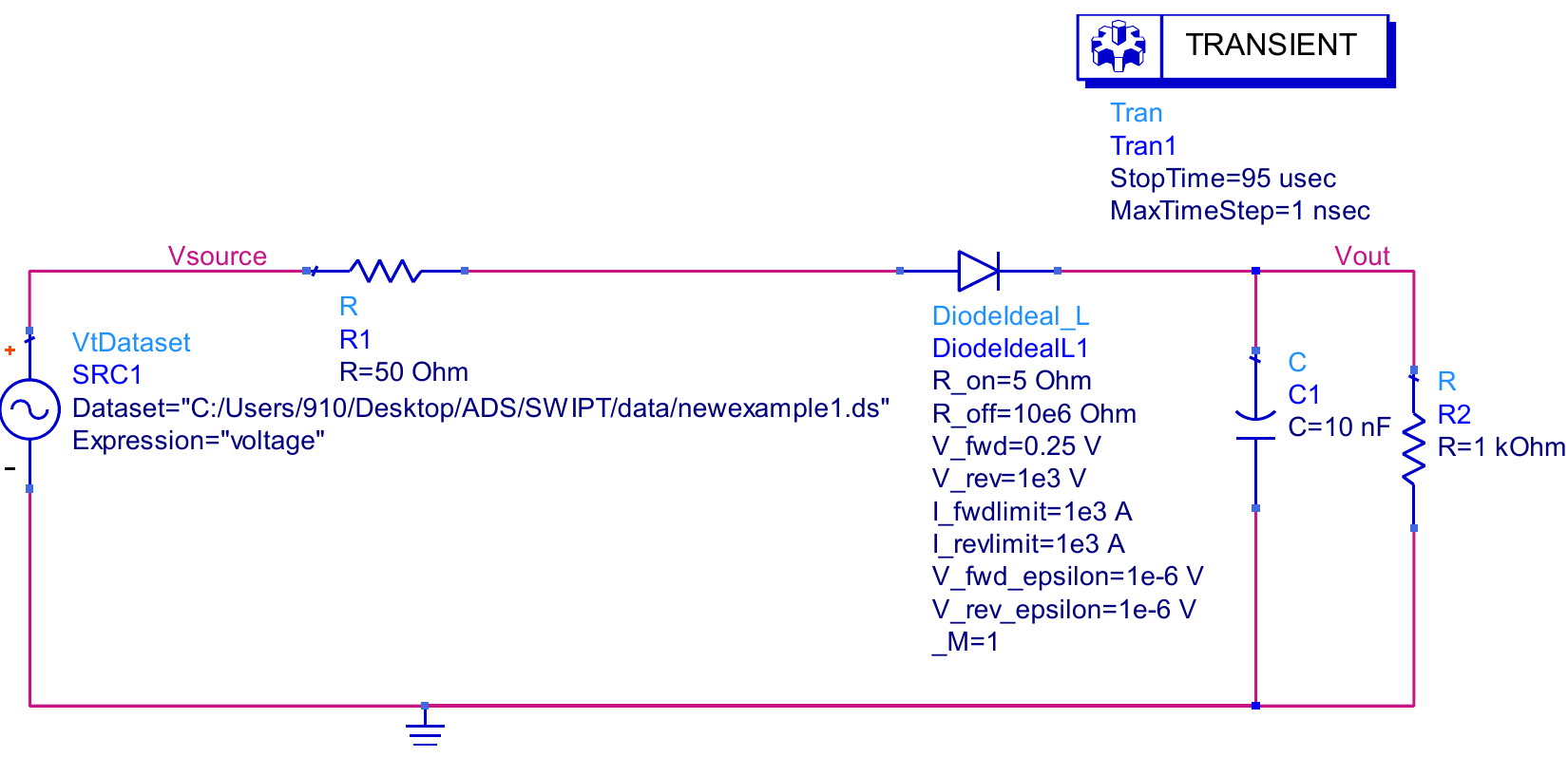}
\caption{Half-wave rectifier on ADS.}\label{ADScircuit}
\vspace{-3mm}
\end{figure}

In order to capture the rectifier's output within a time period we make use of the two expressions in \eqref{Don} and \eqref{Doff} along with the conditions for the diode to switch on (forward biased) or off (reverse biased). The circuit's output is then obtained through a transient time analysis. The pseudo code used for carrying out the transient time output of the circuit is depicted in Algorithm \ref{A1}, where we consider $100$ samples per input period \cite{thesis}. Moreover, besides obtaining the analytical output, the half-wave rectifier is implemented and evaluated through simulations by using the ADS software as depicted in Fig. \ref{ADScircuit}. 

\begin{figure}[t]\centering
\includegraphics[width=0.9\linewidth]{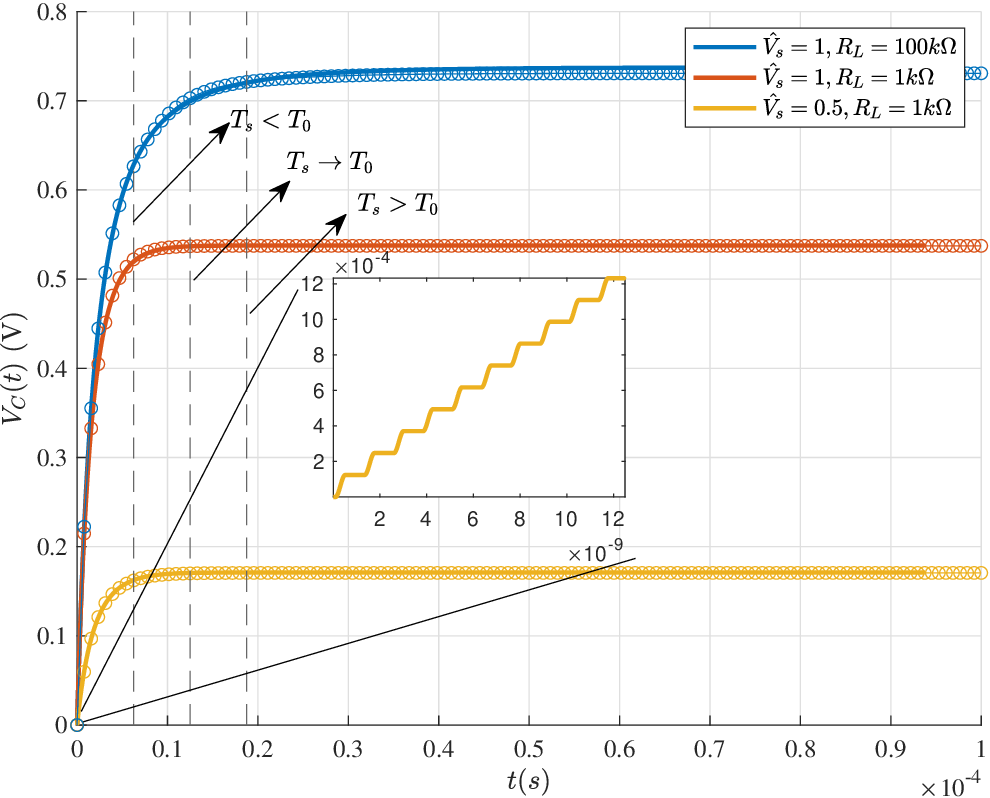}
\caption{Rectifier's output voltage $V_C(t)$; Lines and markers correspond to analytical and ADS simulation, respectively; $f=800$ MHz ($\omega=2\pi f$), $C=10$ nF, $\ron=5 \Omega$, $R_s=50$ $\Omega$, $\roff=10$ M$\Omega$, $\von=0.25$ V and $\h=1$.}\label{nchannel}
\vspace{-5mm}
\end{figure}

\begin{algorithm}[t]
\caption{Rectifier's transient time output}\label{A1}
\begin{algorithmic}[1]
\State $state=0$  \Comment{Diode is off}
\State $\vout(t)=0$ \Comment{Output voltage is zero}
\State $\toff=0$ \Comment{Set initial conditions}
\State set $V_0^{\rm off}=0$
\For{$t\leq T_{sim}$}  \Comment{$T_{sim}$=Total evaluation time}
\\$V_s(t)=\h \sin(\w t)$
\If{$state = 0$}
\State $\vout(t)=$ Eq. \eqref{Doff}
\If{$V_D(t) \geq \von$}
\State $state=1$ 	\Comment{Diode turns on}
\State set $\ton=t$ \Comment{Update initial conditions}
\State set $V_0^{\rm on}=\vout(t)$
\EndIf
\ElsIf{$state = 1$}
\State $\vout(t)=$ Eq. \eqref{Don}
\If{$V_D(t) < \von$}
\State $state=0$ 	\Comment{Diode turns off}
\State set $\toff=t$ \Comment{Update initial conditions}
\State set $V_0^{\rm off}=\vout(t)$
\EndIf
\EndIf
\EndFor
\end{algorithmic}
\end{algorithm}
The output of the algorithm and simulation from ADS is illustrated in Fig. \ref{nchannel}, for constant $\h$, revealing that the analytical results are matched with the simulation results. In addition, we can observe that the output voltage across the load increases in each period. Moreover, over a long-time period the output of the circuit is captured by an exponential curve reaching a certain steady state. In reality though, the output voltage increases gradually within each period in a non-continuous manner as depicted within the zoomed figure. Also, when the load of the circuit is very large and acts as an open circuit, the capacitor's charge increases rapidly until the voltage across reaches $V_C(t)\approx\h-\von$. On the other hand, when a load is present at the circuit, the capacitor discharges accordingly and the achieved steady state is at a lower value of $V_C(t)$.
\section{Memory-based Integrated SWIPT Receiver}
In this section, we apply the presented half-wave rectifier model to a simple communication setup with an integrated SWIPT receiver. Specifically, by considering amplitude modulation, we investigate the circuit's ability for symbol detection based on its output. The motivation for using amplitude modulation stems from the fact that the rectifier converts the input to one direction current flow \cite{SofieC}. Now, in order to achieve a non-zero harvested power, we consider the following: \begin{itemize}
    \item[($i$)] A biased $M$-ASK modulation scheme is applied \cite{SofieC}, \cite{SofieJ}, where the $k$-th information symbol is mapped to a certain amplitude $\h=A_k$, $k\in \{0, 1, \dots, M-1\}$.
    \item[($ii$)] A minimum symbol amplitude $A_k\geq A_{min}> \von$ is set since the charging period of the capacitor starts once the diode's threshold is surpassed.
\end{itemize}

At the receiver's side, we assume that sampling is performed at the end of each symbol period. By denoting the symbol period by $T_s$, the $i$-th symbol can be written as
\begin{equation}
x_i = V_C(t-iT_s).\label{x}
\end{equation}
We assume that the noise due to sampling is an AWGN and thus the $i$-th symbol to be detected is
\begin{align}
y_i &= x_i+n_i\nonumber\\
&= V_C(t-iT_s)+n_i,\label{y}
\end{align}
where $n_i \sim \mathcal{N}\left(0,\sigma^2\right)$ and $\sigma^2$ denotes the variance of AWGN.

As explained in Section \ref{TTA}, under specific circuit parameters, for each symbol amplitude at the input of the circuit i.e., $V_s(t)=A_k \sin(\w t)$, the capacitor achieves a specific steady state after a certain time period. Let the steady state achieved with symbol $A_k$ be denoted by $a_k$. Thus, for $A_k > A_l$, it holds that $a_k > a_l$, where $k\neq l$. Note that, with symbol $A_k$ as the input at the circuit, the capacitor will charge or discharge accordingly from its current state towards reaching $a_k$. Moreover, to mitigate intersymbol interference, the symbol amplitudes are chosen in such a way so that $||A_k-A_l||$ is large. This results in $||a_k-a_l||$ also being large, which corresponds to the largest euclidean distance that any two symbols would have between them.

In what follows, we provide two detection schemes that can be applied based on the symbol period $T_s$. More precisely, if $T_s$ is large enough to lead the capacitor to its steady state, then we know that $x_k \approx a_k$. On the other hand, the value of $x_k$ cannot be determined accurately (it depends on the capacitor's memory) and thus there is a higher probability of error. Clearly, the duration needed for reaching the steady state from an arbitrary level varies. However, for the sake of simplicity and brevity, we consider the cases $T_s > T_0$ and $T_s < T_0$, where $T_0$ is the time interval needed for reaching the steady state from an empty capacitor; for instance, in Fig. \ref{nchannel}, $T_0 \approx 15$ $\mu$s for the case $R_L = 1$ k$\Omega$.

\subsection{Maximum-Likelihood Detection}
We first consider the conventional symbol-by-symbol ML detection scheme. It is worth mentioning that, the ML detector is always applicable in the case $T_s  > T_0$, which refers to an interference-free scenario. Specifically, when $T_s> T_0$, then $x_k \approx a_k$ and the $i$-th detected symbol $\hat{x_i}$ is expressed by
\begin{equation}
\hat{x_i}=\arg \min_{x_j} |y_i-x_j|,
\end{equation}
where $y_i$ and $x_j$ are given by \eqref{y} and \eqref{x}, respectively. Let us now focus on a smaller symbol period i.e., $T_s < T_0$. Even though a larger information rate is achieved, a smaller symbol period does not guarantee that the circuit reaches a steady state. In fact, for a given symbol period smaller than $T_0$, it is not possible to determine where exactly the level of the capacitor is. However, it is possible to provide a range for the output values for each symbol input. In the case where these ranges do not overlap, we can exploit them to form a decision rule and provide an upper bound for the theoretical performance. Specifically, let $r_k^{\min}$ and $r_k^{\max}$ denote the minimum and maximum output, respectively, of the $k$-th symbol when adjacent symbols are transmitted; these can be obtained with Algorithm \ref{A1}. Then, the decision region for the $k$-th symbol is $\rho_1 \leq y_i \leq \rho_2$, where $\rho_1 = (r_k^{\min}-r_{k-1}^{\max})/2$ and $\rho_2 = (r_{k+1}^{\min}-r_k^{\max})/2$. On the other hand, if these ranges overlap, the symbols cannot be distinguished between them. This effect can be avoided through the proper design of the circuit, that is, through its parameters such as minimum input, maximum transmitted power, etc. This allows a constellation with sufficiently large euclidean distance between the symbols such that the decision regions are non-overlapped.
\begin{figure}[t]\centering
\includegraphics[width=0.9\linewidth]{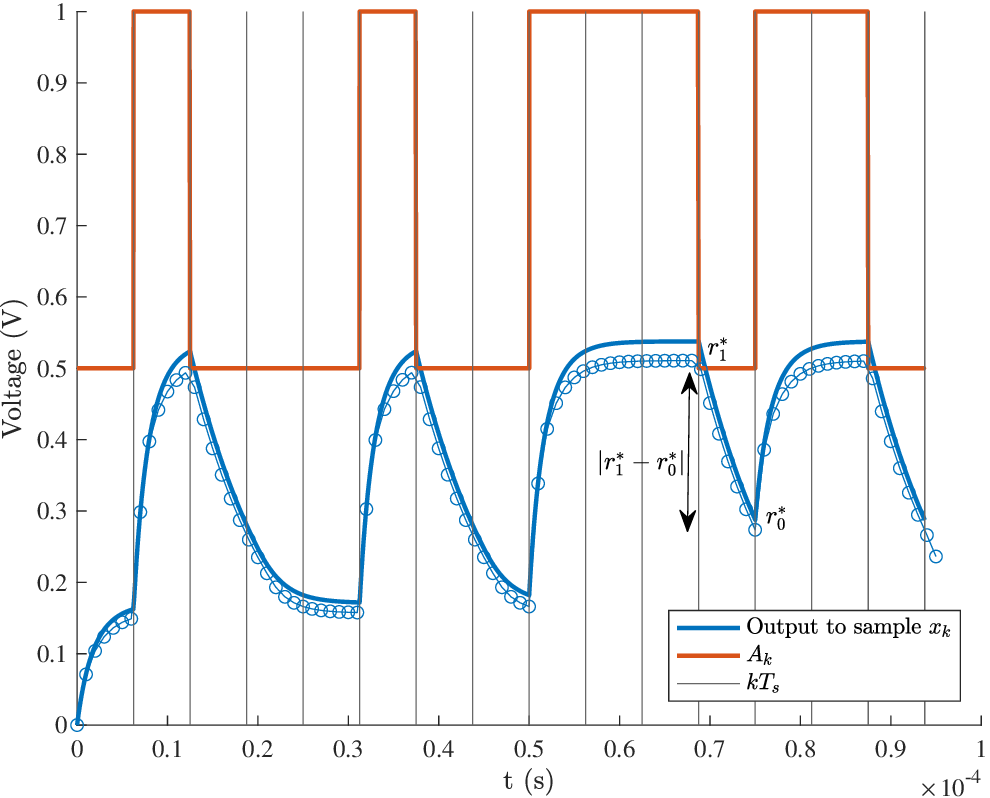}
\caption{Transmission of $15$ bits with $T_s=6.25\mu s$; markers correspond to ADS simulation; $f=800$ MHz, $C=10$ nF, $R_L=1k$ $\Omega$, $R_s=50$ $\Omega$, $\ron=5 \Omega$, $\roff=10$ M$\Omega$, $\von=0.25$.}\label{15bits}
\vspace{-5mm}
\end{figure}

In what follows, we provide a specific example for the binary ASK (BASK) modulation. When $T_s > T_0$, the decision bound for detecting symbol $A_1$ is $y_i\geq \rho$ and for $A_0$ is $y_i < \rho$, where $\rho = |a_0+a_1|/2$. Then, since the two symbols are equiprobable, the error probability over AWGN is given by
\begin{align}
\mathcal{P}_e = Q\left(\frac{a_1-a_0}{2\sigma}\right),
\end{align}
where $Q(\cdot)$ is the $Q$-function. Now, when $T_s < T_0$, the rectifier's output alternates in specific ranges for each symbol. 
Let $r_1^*$ and $r_0^*$, denote the minimum and maximum output of symbols with mapping amplitudes $A_1$ and $A_0$. These output values would result in the minimum euclidean distance between the sampled symbols i.e., $\min\{||x_1-x_0||\}=||r_1^*-r_0^*||$. An example is depicted in Fig. \ref{15bits}, indicating the closest values of two consecutive and adjacent symbols with $A_1=1$ and $A_0=0.5$. In this example, $T_s< T_0$ and we can observe the symbol duration relative to the steady state in Fig. \ref{nchannel}. Following from the aforementioned, for the symbol detection, ML can be applied by assuming that $x_j=r_j^*$, where $j \in \{0,1\}$. Recall that, as the symbol period increases, then $x_j \to a_j$ and $r_j^* \to a_j$. In this case, the theoretical upper bound captures the actual error probability. 

\subsection{Maximum Likelihood Sequence Detection}
A general detection scheme which is optimal for any symbol period $T_s$, is the maximum likelihood sequence detection. In this case, the input at the detector consists of $K$ received symbols forming a sequence of length $K$ which is denoted by $\mathbf{y}$. Then, a sequence is detected according to 
 \begin{equation}
 \hat{\mathbf{x}}=\arg \min_{\mathbf{x}_m} ||\mathbf{x}_m-\mathbf{y}||^2,
 \end{equation}
where $\mathbf{x}_m$ denotes the $m$-th sequence of $K$ consecutive symbols. In order to obtain all the possible sequences, transient time analysis of the circuit is used over $M^K$ sequences. Note that, since the capacitor induces memory to the system $M^K$ sequences should be obtained for all the possible output levels. For ease of comparison, we assume that the system has a finite memory of $L$ symbols. This corresponds to sequential transmission of $L$ symbols followed by an artificial replacement of the charged capacitor with an empty capacitor. Note that, the sequence detection becomes necessary when the symbol period is small relative to reaching the steady state due to the memory effects of the capacitor.

\subsection{Energy Harvesting Performance}
In order to capture the energy harvesting performance we measure the instantaneous power at the load at the end of each symbol period. The instantaneous power at the load is given by 
\begin{equation}
P_l(t)=\frac{V_C(t)^2}{R_l}.
\end{equation}
For a sequential transmission of length $L$ we evaluate the average power of a sequence by
\begin{equation}
P_L=\frac{1}{L}\sum_{i=1}^{L} P_l(t-iT_s).
\end{equation}

\section{Numerical results}
\begin{figure}[t]\centering
\includegraphics[width=0.9\linewidth]{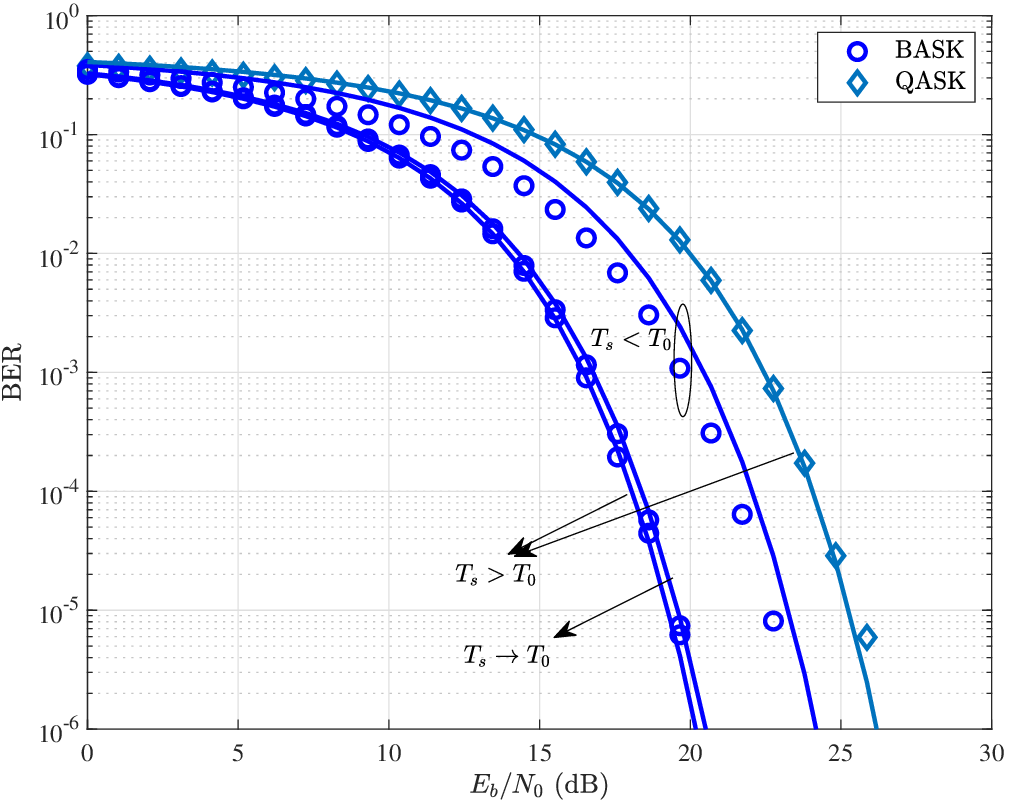}
\caption{Bit error rate performance with respect to SNR with symbol-by-symbol ML detection.}\label{BERmine}
\end{figure}

\begin{figure}[t]\centering
\includegraphics[width=0.9\linewidth]{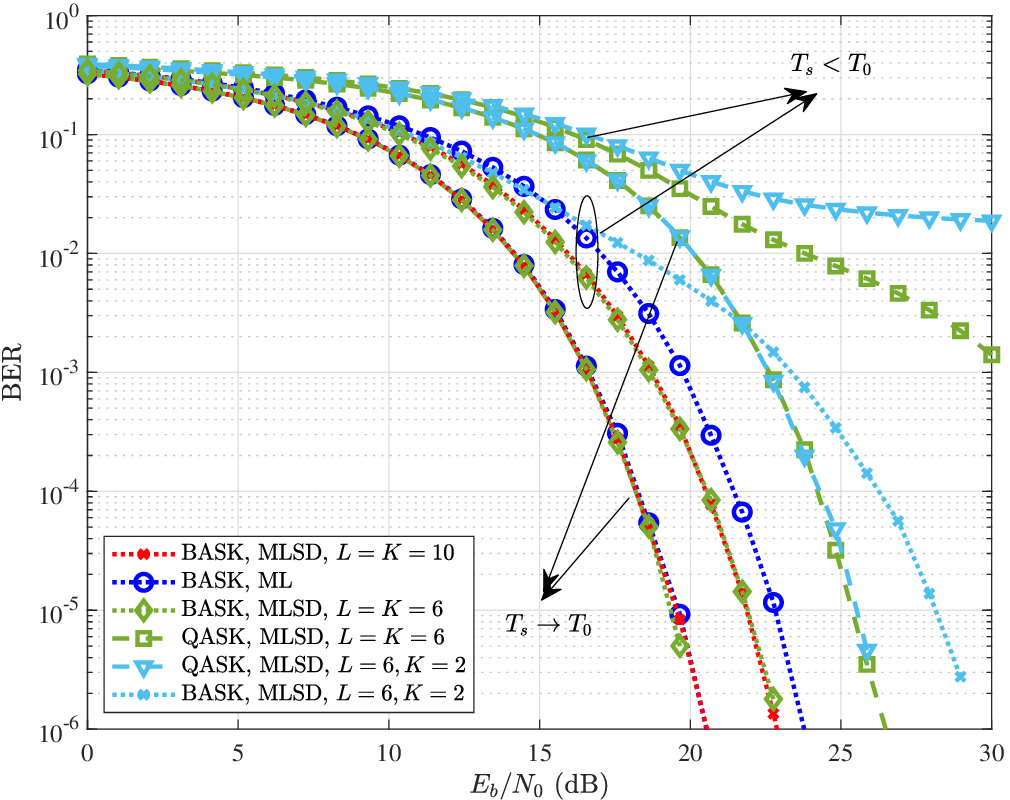}
\caption{Bit error rate performance with respect to SNR with MLSD.}\label{seq}
\vspace{-5mm}
\end{figure}
In this section we provide the SWIPT receiver's performance in terms of both, ID and EH. For this purpose we use transient time analysis as explained in Section \ref{TTA}. Unless otherwise stated, we use  $f=800$ MHz ($\omega=2\pi f$), $C=10$ nF, $R_L=1k$ $\Omega$, $R_s=50$ $\Omega$, $\ron=5 \Omega$, $\roff=10$ M$\Omega$, $\von=0.25$. For the BASK modulation scheme we consider $A_{min}=0.5$ hence the amplitudes $A_0=0.5$ and $A_1=1$ are used. The normalized $\sigma$ is obtained based on the transmitted symbol power as follows. The $m-$th symbol's power is given by $P_m=\frac{1}{2T_s}\int_{-T_s}^{T_s}|A_m\sin\left(\omega_0t\right)|^2\,dt={A_m^2}/{2}$, and the average symbol power is given by $P_{av}=\frac{1}{4}\sum_{m=0}^{1}A_m^2$. Then, since $\frac{E_b}{N_0}=\frac{P_{av}}{2 \sigma^2}$, it occurs that $\sigma=\sqrt{\frac{P_{av}}{2 E_b/N_0}}$, where in this case $P_{av}=\frac{5}{16}$. Moreover, for a fair comparison with the quadrature ASK (QASK) modulation, we keep the average symbol power constant as well as the same minimum symbol amplitude i.e., $A_{min}=0.5$ and set $A_m=A_{min}+m d_s$, where $d_s=(1/14)(\sqrt{30}-3)$ and $m\in\{0,\dots,3\}$. The corresponding normalized $\sigma$ in this case is given by $\sigma=\sqrt{\frac{P_{av}}{4 E_b/N_0}}$, since two bits per symbol are transmitted. For the sake of comparison we consider three symbol periods, $T_s< T_0=6.25$$\mu s$ and $T_s\to T_0=12.5$$\mu s$ and $T_s >T_0=18.75$$\mu s$ which are indicated in Fig. \ref{nchannel} to depict the symbol period's selection with respect to the steady state output.

In Fig. \ref{BERmine}, we plot the BER performance based on symbol-by-symbol ML detection with respect to the signal to noise ratio per bit $E_b/N_0$. We plot BASK and QASK for the case where $T_s>T_0$, i.e., each symbol is sampled at its corresponding steady state. As can be seen, in both cases the transient time analysis matches with the theoretical error probability obtained by utilizing the steady state values. Furthermore, for the BASK modulation we compare the performance over three different symbol periods. For the cases where $T_s\leq T_0$, in order to define the decision regions for each symbol, we first obtain $r_0^*$ and $r_1^*$ which are then used for the defining the symbols' detection regions. Clearly, when a smaller symbol period is used, a worse performance occurs as the boundary between the symbols becomes smaller. This is due to the fact that there is not enough time for the capacitor to reach the corresponding symbol's steady state. Moreover, the theoretical $\mathcal{P}_e$, in these cases correspond to an upper bound as it captures the closest euclidean distance it might occur. As $T_s$ increases $r_k^*\approx a_k$ and the theoretical $\mathcal{P}_e$ captures the actual performance which is better compared to applying a smaller $T_s$, since a larger euclidean distance between the symbols occurs. For the QASK modulation, a smaller $T_s$ results in a more severe intersymbol interference which imposes the use of MLSD in order to achieve higher data rate.

In Fig. \ref{seq}, we consider sequential transmission i.e., a finite memory of length $L=6$ symbols. We plot the BER with respect to $E_b/N_0$ for BASK and QASK where MLSD is applied with $T_s\in\{6.25, 12.5\}$ $\mu s$. As expected, when the symbol period and the sequence length $K$, increase, a better performance is achieved. Moreover, for the BASK we also compare the MLSD performance with the symbol-by-symbol ML. Clearly, when $T_s<T_0$, the MLSD detection outperforms ML. However, when $T_s\to T_0$, the MLSD converges to the ML detection since the effect of the memory in the system is eliminated and there is no intersymbol interference. Moreover, when $L=K$, the performance is the same even for higher $L$ i.e., the same performance is achieved when $L=10$. 

\begin{figure}[t]\centering
\includegraphics[width=0.8\linewidth]{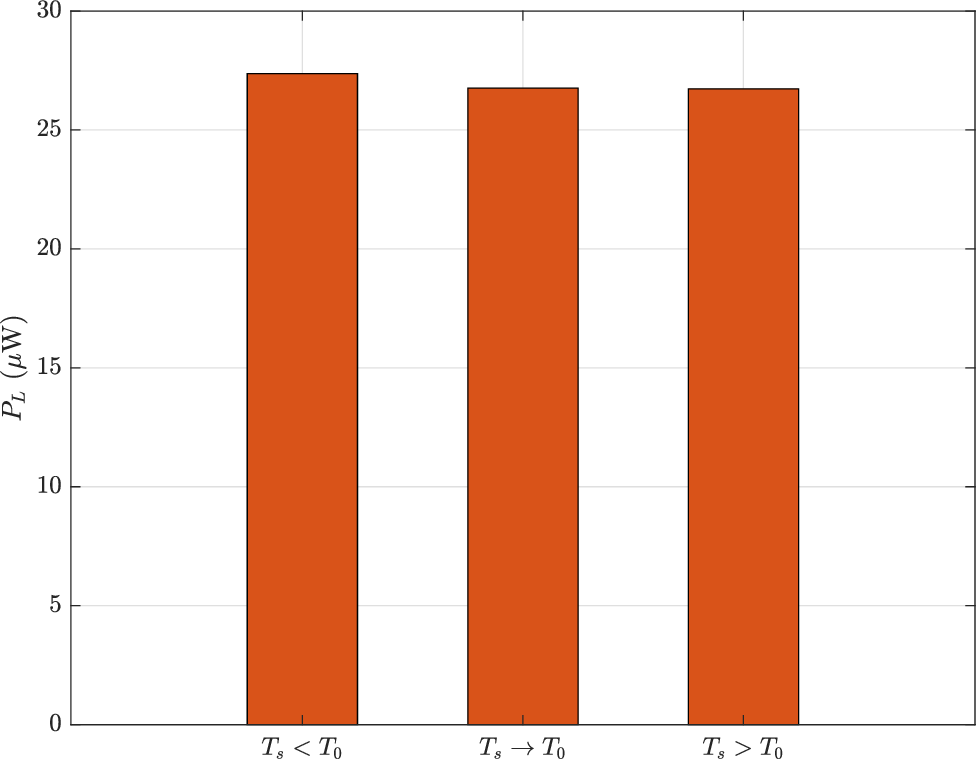}
\caption{Average sequence power for BASK, $L=6$.}\label{EH}
\end{figure}

Finally in Fig. \ref{EH}, we plot the average sequence output power for three different symbol periods when BASK is applied. Interestingly, the best performance is achieved when a lower $T_s$ is used. This occurs from the fact that a lower symbol period does not allow the capacitor to discharge enough towards the lower steady state that a low amplitude symbol would result.  

\section{Conclusions}
This paper proposed an accurate and simple-to-use model to characterize the output voltage of an EH circuit based on the half-wave rectifier topology. The validity of the analytical expressions was confirmed with ADS simulations. The proposed model was further explored in the context of a SWIPT scenario under the memory effects introduced by the capacitor. We showed that, with ML detection, the symbol period is critical to the BER performance due to intersymbol interference. Therefore, we also applied MLSD, which improves the BER and allows for high data rates.

\end{document}